\documentstyle[12pt]{article}
\long\def\comment#1{}
\begin{document}
\title{Concepts of Space, Time, and Consciousness in Ancient India}
\author{Subhash Kak\\
Department of Electrical \& Computer Engineering\\
Louisiana State University\\
Baton Rouge, LA 70803-5901\\}
\maketitle

\begin{abstract}
This paper describes Indian ideas of the 
early-Pur\={a}\d{n}a/Mah\={a}bh\={a}rata times
(centuries BC) on
the nature of space, time and consciousness that
would be of interest to the physicist.
In order to simplify references, we quote mainly from 
Yoga-V\={a}si\d{s}\d{t}ha (YV), which is representative
of that period of Indian thought.
YV professes to be a book of instruction 
on the nature of consciousness but it has
many fascinating passages on time, space, matter and
cognition.
This paper presents a random selection that has parallels with
recent speculations in physics.
It also presents a brief account of the context in which ideas of
YV developed.
\end{abstract}

\section{Introduction}
Ancient Indian ideas of physics, available to us
through  a variety of sources, are generally not
known in the physics world.
Indian astronomer/physicists, starting with
a position that sought to unify space, time,
matter, and consciousness,
argued for relativity of space and time, cyclic
and recursively defined
universes, and a non-anthropocentric view.
The two most astonishing numerical claims from the
ancient Indians are: a cyclic system 
of creation of the universe with a period of
8.64 billion years, although there exist
longer cycles as well; and, speed of light
to be 4,404 yojanas per nime\d{s}a, which is
almost exactly 186,000 miles per second (Kak, 1998a)!

A critic would see the numbers as no more
than idle coincidences. But within the Indian tradition
it is believed that reality, as a kind of a
{\em universal state function}, transcends the separate
categories of space, time, matter, and observation.
In this function, called {\em Brahman} in the literature,
inhere all categories including
knowledge. 
The conditioned mind can, by ``tuning'' in to {\em Brahman},
obtain knowledge, although it can only be expressed
in terms of the associations already experienced by
the mind.
Within the Indian tradition, scientific knowledge
describes as much aspects of outer reality as
the topography of the mindscape.
Furthermore, there are connections between the outer
and the inner: we can comprehend reality only because
we are already equipped to do so!
My own papers listed in the bibliography can serve as
an introduction to these ideas and point
to further references for the reader to examine.

Two philosophical systems at the basis of
Indian physics---and
metaphysics---are S\={a}\d{m}khya and Vai\'{s}e\d{s}ika.
S\={a}\d{m}khya, which is an ancient
system that goes back to the 3rd millennium BC, 
posits 25 basic categories together with 3 constituent
qualities, which evolve in different ways to
create the universe at the microcosmic as well
as the macrocosmic levels.
It also presupposes a ``potential'' ({\em tanm\={a}tra}) to be more
basic than the material entity.
Vai\'{s}e\d{s}ika
is a later system which
is an atomic theory with the non-atomic
ground of ether, space, and time upon which rest
four different classes of indestructible atoms which
combine in a variety of ways to constitute all matter;
it also considers mind to be atomic (Kak, 1999).
These systems presuppose genesis and evolution both at the cosmic
and psychological levels.
They also accept cyclic and multiple universes, and centrality
of observers.

Unfortunately, historians of science are generally
oblivious of Indian physics, astronomy or cosmology.
Amongst popular books,
Paul Halpern's {\em The Cyclical Serpent} (1995) is 
unusual in that it places modern speculations
regarding an oscillating universe within the
context of the cyclic cosmology of the Pur\={a}\d{n}as,
but even this book doesn't define a context for the
Indian ideas.

In this paper we present, in a capsule form, the
basic Indian ideas on space, time, and
observation from the age of the epics and
the early Pur\={a}\d{n}as.
The ideas of these period seem to belong
to last centuries BC and they are described
in the Mah\={a}bh\={a}rata, Pur\={a}\d{n}as,
and the early Siddh\={a}ntas.
To keep our sources to a minimum, we mainly use 
Yoga-V\={a}si\d{s}\d{t}ha (YV), an ancient Indian text,
over 29,000 verses long,  traditionally
attributed to V\={a}lm\={\i}ki, author of the epic R\={a}m\={a}ya\d{n}a, which
is over two thousand years old.

\section{Vedic and Pur\={a}\d{n}ic Cosmology}

We first look at Vedic cosmology. The Vedas are
texts that represent the ancient knowledge 
tradition of India. While their compilations go back to at
least the third millennium, some of their
contents might be even older.
The Vedic tradition is a part of the Indian culture
tradition that has been traced back, archaeologically, to about
8000 BC (Feuerstein et al, 1995).
The antiquity of the Vedic texts is, in part, confirmed by
their celebration of the Sarasvati river as the greatest
river of their age, and modern hydrological studies have
established that this river dried up around 2000 BC.
The king-lists in the Pur\={a}\d{n}as take us to
several millennia before the period of the drying up
of the Sarasvati.
There is also a rock art tradition in India that has been
traced to about 40000 BC (Wakankar, 1992).

There are several statements in the Vedic texts about
the universe being infinite, while at the same time
the finite distance to the sun is explicitly
mentioned (Kak, 1998a-d).
Aditi, the great mother of the gods, is a personification of the
concept of infinity.
A famous mantra speaks of how taking infinity out of infinity
leaves it unchanged.
This indicates that paradoxical properties of the notion of
infinity were known.

In a reference to mapping the outer world into
an altar made of bricks,
the Yajurveda (hymn 17) names numbers
in multiples of ten that go upto ten hundred thousand million.
This also suggests a belief in a very large universe.

The \'{S}atapatha Br\={a}hma\d{n}a,
a commentatorial prose text on the 
Veda, that most likely goes back to the early centuries
of the second millennium BC,
provides
an overview of some broad aspects of Vedic cosmology.
The sixth chapter of the book, entitled
``Creation of the Universe'', speaks of
the creation of the earth later than that of
other stars.
Creation is seen to proceed
under the aegis of the Praj\={a}pati (reference either to a star
or to abstract time)
with the emergence of
A\'{s}va, R\={a}sabha, Aja and K\={u}rma before the emergence of the
earth.
Vi\'{s}van\={a}tha Vidy\={a}la\.{n}k\={a}ra suggests that these
are the sun (A\'{s}va),
Gemini (R\={a}sabha), Aja (Capricorn) and
K\={u}rma (Cassiopeia).
This identification is supported by
etymological considerations.
The \d{R}gveda 1.164.2 and Nirukta 4.4.27 define A\'{s}va as the sun.
R\={a}sabha which literally means the twin asses
are defined in Nighan\d{t}u 1.15 as A\'{s}vinau which later usage
suggests are
Castor and Pollux in Gemini.
In Western astronomy the twin asses are to be found in the next
constellation of Cancer as Asellus Borealis and Asellus
Australis.
Aja (goat) is defined by Nighan\d{t}u 1.15 as a sun and owing to
the continuity that we see in the Vedic and later European names
for constellations (as in the case of the Great Bear) it is
reasonable to identify it as the constellation Capricorn
({\em caper} goat + {\em cornu} horn).
K\={u}rma is a synonym of Ka\'{s}yapa (tortoise) which is like
Cassiopeia (from Greek Kassiopeia),
and it is appropriate because it is near the pole.

The Pur\={a}\d{n}as view the universe to have
a diameter of about 500 million yojanas, but beyond
the universe lies the limitless {\it Pradh\={a}na},
that has within it countless other universes (Kak, 1998a).

\section{The Yoga-V\={a}si\d{s}\d{t}ha}
The internal evidence of the 
Yoga-V\={a}si\d{s}\d{t}ha (YV) 
indicates that it was authored or
compiled later than the 
R\={a}m\={a}ya\d{n}a.
Chapple (1984) summarizes the views of various scholars who date it
variously as early as the sixth century AD or as late as
the 13th or the 14th century.
Dasgupta (1975, 1932) dated it about the sixth century AD on the basis
that one of its verses appears to be copied from one
of K\={a}lid\={a}sa's
plays considering 
K\={a}lid\={a}sa to have lived around the fifth century. 
The traditional date
of K\={a}lid\={a}sa is 50 BC and 
new arguments
(Kak 1990) support this earlier date so that 
the estimates regarding the age of
YV are further muddled and it is possible
that this text could be 2000 years old.

YV may be viewed as a book of philosophy or as a philosophical novel.
It describes the instruction given by 
Vasi\d{s}\d{t}ha to R\={a}ma, the hero of the epic
R\={a}m\={a}ya\d{n}a.
Its premise may be termed radical idealism and it is couched in
a fashion that has many parallels with the notion of a
participatory universe argued by Wheeler and others.
Its most interesting passages from the scientific point of view
relate to the description of the nature of space, time, matter, and
consciousness.
It should be emphasized that the YV ideas do not stand in isolation.
Similar ideas are to be found in the earlier Vedic books.
At its deepest level the Vedic conception is to view reality in a monist
manner; at the next level one may speak of the dichotomy of mind and matter.
Ideas similar to those found in YV are also encountered in Pur\={a}\d{n}as
and Tantric literature.

YV is a text that belongs to the mainstream of the ancient Vedic
tradition that professes to deal with knowledge.
Astronomical references in the Vedic texts take us back to the 4th
or 5th millennium BC or even earlier (e.g. Kak 1994-6).

Roughly speaking, the Vedic system speaks of an interconnectedness between
the observer and the observed. A similar conception appears to have informed
many ancient peoples including the Greeks. 

The Vedic system of knowledge is based on a tripartite approach to the
universe where connections exist in triples in categories of
one group and across groups:
sky, atmosphere, earth; object, medium, subject; future, present, past;
and so on. Beyond the triples lies the transcendental ``fourth''.

Three kinds of motion are alluded to in the Vedic books:
these are the translational motion, sound, and light which 
are taken to be ``equivalent'' to earth, air, and sky.
The fourth motion is assigned to
consciousness; and this is considered to be infinite in speed.

At least one of the founders of quantum theory was directly inspired
by the Vedic system of knowledge.
Schr\"{o}dinger (1961) claims that the Vedic slogan ``All in One and One
in All'' was an idea that led him to the creation of quantum mechanics
(see also Moore, 1989).
Even before Schr\"{o}dinger, the idealist philosophical tradition in
Europe had long been moulded by Vedic ideas.
It should also be noted that many parts of the Vedic literature are still
not properly understood although considerable progress has recently
taken place in the study of Vedic science.

It is most interesting that the books in this Indian tradition speak about the
relativity of time and space in a variety of ways.
The medieval books call the Pur\={a}\d{n}as speak of countless universes,
time flowing at different rates for different observers and so on.

\comment{
There is an ancient mention of space travelers wearing airtight suits
in the epic Mah\={a}bh\={a}rata which may be classified as an early form
of science fiction. According to the well-known Sanskritist J.A.B.
van Buitenen, in the accounts in Book 3 called ``The Razing of Saubha''
and ``The War of the Yak\d{s}as'':

\begin{quote}
the aerial city is nothing but an armed camp with flame-throwers and
thundering cannon, no doubt a spaceship. The name of the demons is also
revealing: they were Niv\={a}takavacas, ``clad in airtight armor,'' which
can hardly be anything but space suits. (van Buitenen, 1975, page 202)
\end{quote}
 
The context of modern science fiction books is clear: it is the liberation
of the earlier modes of thought by the revolutionary developments of
the 20th century science and technology.
How was science fiction integrated into the
mainstream of Indian literary tradition two thousand years ago is not
properly understood.}
 
Universes defined recursively are described in
the famous episode of Indra and the ants in Brahmavaivarta Pur\={a}\d{n}a
4.47.100-160, the Mah\={a}bh\={a}rata 12.187, and elsewhere.
These flights of imagination are to be traced to more than a straightforward
generalization of the motions of the planets into a cyclic universe.
They must be viewed in the background of an amazingly sophisticated
tradition of cognitive and analytical thought (see e.g. Staal 1988;
Rao and Kak 1998).

\subsection*{Selected Passages}
The page numbers given at the end of each passage are from the
Venkatesananda (1993) translation. YV consists of 6 books where
the sixth book itself has two parts. The numbers in the square
brackets refer to the book, (part), section, verse.
The reference to the Sanskrit original is also listed in
the bibliography.

\subsection*{Time}
\begin{itemize}

\item Time cannot be analyzed; for however much it is divided
it survives indestructible. [1.23]

\item There is another aspect of this time, the end of
action ({\em k\d{r}t\={a}nta}), according to the law of nature
({\em niyati}). [1.25.6-7]
\item The world is like a potter's wheel: the wheel looks as if it
stands still, though it revolves at a terrific speed. [1.27]
\item Just as space does not have a fixed span, time does not have
a fixed span either. Just as the world and its creation are mere
appearances, a moment and an epoch are also imaginary. [3.20]
\item Infinite consciousness held in itself the notion of a unit of
time equal to one-millionth of the twinkling of an eye: and from this
evolved the time-scale right upto an epoch consisting of several
revolutions  of the four ages, which is the life-span of one
cosmic creation. Infinite consciousness itself is uninvolved in
these, for it is devoid of rising and setting (which are essential to
all time-scales), and it devoid of a beginning, middle and end. [3.61]
\end{itemize}

\subsection*{Space}
\begin{itemize}

\item There are three types of space---the psychological space, the
physical space and the infinite space of consciousness. [3.17]

The infinite space of individed consciousness is that which exists in all,
inside and outside...
The finite space of divided consciousness is that which created divisions of
time, which pervades all beings...
The physical space is that in which the elements exist.
The latter two are not independent of the first. [3.97]
\item {\it Other universes/wormholes.} I saw within [the] rock [at the
edge of the universe] the creation,
sustenance and the dissolution of the universe... I saw innumerable
creations in the very many rocks that I found on the hill.
In some of these creation was just beginning, others were
populated by humans, still others were far ahead in the
passage of their times. [6.2.86]
\item I perceived within each molecule of air a whole
universe. [6.2.92]
\end{itemize}

\subsection*{Matter}
\begin{itemize}

\item In every atom there are worlds within worlds. [3.20]

\item I saw reflected in that consciousness the image of
countless universes.
I saw countless creations though they did not know of one
another's existence.
Some were coming into being, others were perishing, all
of them had different shielding atmospheres (from five to
thirty-six atmospheres).
There were different elements in each, they were inhabited
by different types of beings in different stages of evolution..
[In] some there was apparent natural order in others there
was utter disorder, in some there was no light and
hence no time-sense. [6.2.59]
\end{itemize}

\subsection*{Experience}
\begin{itemize}

\item Direct experience alone is the basis for all proofs...
That substratum is the experiencing intelligence which itself becomes
the experiencer, the act of experiencing, and the experience. [2.19-20]
\item Everyone has two bodies, the one physical and the other mental.
The physical body is insentient and seeks its own destruction; the mind
is finite but orderly. [4.10]
\item I have carefully investigated, I have observed everything from the tips
of my toes to the top of my head, and I have not found anything of which I
could say, `This I am.' Who is `I'?
I am the all-pervading consciousness which is itself not an object of
knowledge or knowing and is free from self-hood. I am that which is
indivisible, which has no name, which does not undergo change, which is
beyond all concepts of unity and diversity, which is beyond measure. [5.52]
\item I remember that once upon a time there was nothing on this earth,
neither trees and plants, nor even mountains.
For a period of eleven thousand years the earth was covered by lava. In
those days there was neither day nor night below the polar region: for in the
rest of the earth neither the sun nor the moon shone. Only one half of the polar
region was illumined.

Then demons ruled the earth. They were deluded, powerful and prosperous,
and the earth was their playground.

Apart from the polar region the rest of the earth was covered with water.
And then for a very long time the whole earth was covered with forests,
except the polar region. Then there arose great mountains, but without
any human inhabitants. For a period of ten thousand years the earth
was covered with the corpses of the demons. [6.1]
\end{itemize}

\subsection*{Mind}
\begin{itemize}

\item The same infinite self conceives within itself the duality of
oneself and the other. [3.1]
\item Thought is mind, there is no distinction between the two. [3.4]
\item The body can neither enjoy nor suffer. It is the mind alone that
experiences. [3.115]
\item The mind has no body, no support and no form; yet by this mind is
everything consumed in this world.
This is indeed a great mystery.
He who says that he is destroyed by the mind which has no
substantiality at all, says in effect that his head was smashed by
the lotus petal...
The hero who is able to destroy a real enemy standing in front of him
is himself destroyed by this mind which is [non-material].
\item The intelligence which is other than self-knowledge is what 
constitutes the mind. [5.14]
\end{itemize}
\subsection*{Complementarity}
\begin{itemize}
\item The absolute alone exists now and for ever. When one thinks of it
as a void, it is because of the feeling one has that it is not void; when one
thinks of it as not-void, it is because there is a feeling that it is
void. [3.10]
\item All fundamental elements continued to act on one another---as
experiencer and experience---and the entire creation came into being
like ripples on the surface of the ocean.
And, they are interwoven and mixed up so effectively that they cannot be
extricated from one another till the cosmic dissolution. [3.12]
\end{itemize}

\subsection*{Consciousness}
\begin{itemize}

\item The entire universe is forever the same as the consciousness that
dwells in every atom, even as an ornament is non-different
from gold. [3.4]
\item The five elements are the seed of which the world is the tree; and
the eternal consciousness is the seed of the elements. [3.13]
\item Cosmic consciousness alone exists now and ever; in it are
no worlds, no created beings. That consciousness reflected in itself
appears to be creation. [3.13]
\item This consciousness is not knowable: when it wishes to become the
knowable, it is known as the universe. Mind, intellect, egotism, the five
great elements, and the world---all these innumerable names and forms are
all consciousness alone. [3.14]
\item The world exists because consciousness is, and the world is the
body of consciousness. There is no division, no difference, no distinction.
{\it Hence the universe can be said to be both real and unreal: real
because of the reality of consciousness which is its own reality, and
unreal because the universe does not exist as universe, independent of
consciousness.} [3.14]
\item Consciousness is pure, eternal and infinite: it does not arise nor
cease to be. It is ever there in the moving and unmoving creatures, in the
sky, on the mountain and in fire and air. [3.55]
\item
Millions of universes appear in the infinite consciousness like specks
of dust in a beam of light. In one small atom all the three worlds appear
to be, with all their components like space, time, action, substance, day
and night. [4.2]
\item The universe exists in infinte consciousness. Infinite consciousness
is unmanifest, though omnipresent, even as space, though existing 
everywhere, is manifest. [4.36]
\item The manifestation of the omnipotence of infinite consciousness enters
into an alliance with time, space and causation. Thence arise infinite
names and forms. [4.42]
\item Rudra is the pure, spontaneous self-experience which is the
one consciousness that dwells in all substances. It is the seed of
all seeds, it is the essence of this world-appearance, it is
the greatest of actions. It is the cause of all causes and it is the
essence of all beings, though in fact it does not cause anything
nor is it the concept of being, and therefore cannot be conceived.
It is the awareness in all that is sentient, it knows itself as its
own object, it is its own supreme object and it is aware of
infinite diversity within itself...

The ifinite consciousness can be compared to the ultimate atom
which yet hides within its heart the greatest of mountains.
It encompasses the span of countless epochs, but it does not let
go of a moment of time.
It is subtler than the tip of single strand of hair, yet it
pervades the entire universe...

It does nothing, yet it has fashioned the universe.
..All substances are non-different from it, yet it
is not a substance; though it is non-substantial it
pervades all substances. The cosmos is its body, yet it has
no body. [6.1.36]

\end{itemize}

\subsection*{The YV model of knowledge}
YV is not written as a systematic text.
Its narrative jumps between various levels:
psychological, biological, and physical.
But since the Indian tradition of knowledge is based on
analogies that are recursive and connect
various domains, one can be certain
that our literal reading of the passages is valid.

YV appears to accept the idea that laws are intrinsic to the universe.
In other words, the laws of nature in an unfolding universe will
also evolve.
According to YV, new information does not emerge out of the inanimate world
but it is a result  of the exchange between mind and matter.

It accepts consciousness as a kind of fundamental field that
pervades the whole universe.
One might speculate that the parallels between YV and some recent ideas
of physics are a result of the inherent structure of the mind.

\section{Other Texts}

Our readings of the YV are confirmed by other texts such as
the 
Mah\={a}bh\={a}rata and the
Pur\={a}\d{n}as as they are by the philosophical systems of
S\={a}\d{m}khya and Vai\'{s}e\d{s}ika, or the various
astronomical texts. 

Here is a reference to the size of the universe from
the 
Mah\={a}bh\={a}rata 12.182:
\begin{quote}
The sky you see above is infinite.
Its limits cannot be ascertained.
The sun and the moon cannot see, above or below,
beyond the range of their own rays.
There where the rays of the sun and the moon cannot reach
are luminaries which are self-effulgent and which
possess splendor like that of the sun or the fire.
Even these last do not behold the limits of the
firmament in consequence of the inaccessibility and
infinity of those limits.
This space which the very gods cannot measure is full
of many blazing and self-luminous worlds each above the
other.\\
(Ganguly translation, vol. 9, page 23)
\end{quote}

The 
Mah\={a}bh\={a}rata has a very interesting passage (12.233),
virtually identical with the corresponding material in YV,
which describes the dissolution of the world.
Briefly, it is stated how a dozen suns burn up
the earth, and how elements get transmuted until
space itself collapses into wind (one of the elements).
Ultimately, everything enters into primeval
consciousness.

If one leaves out the often incongrous commentary on
these ideas which were strange to him, we find 
al-B\={i}r\={u}n\={\i} in his encyclopaedic book
on India written in 1030 speaking of essentially
the same ideas.
Here are two little extracts:
\begin{quote}
The Hindus have divided duration into two periods,
a period of {\em motion}, which has been determined
as {\em time}, and a period of {\em rest}, which
can only be determined in an imaginary way according
to the analogy of that which has first been
determined, the period of motion.
The Hindus hold the eternity of the Creator to be
{\em determinable}, not {\em measurable}, since it
is infinite.

They do not, by the word {\em creation}, understand
a {\em formation of something out of nothing.}
They mean by creation only the working with a piece of
clay, working out various combinations and figures in it,
and making such arrangements with it as will lead to
certain ends and aims which are potentially in it.\\
(Sachau, 1910, vol 1, pages 321-322)
\end{quote}

The mystery of consciousness is a recurring theme
in Indian texts (Kak, 1997). Unfortunately,
the misrepresentation that Indian philosophy is idealistic,
where the physical universe is considered an illusion,
has become very common. For an authoritative
modern exposition of Indian ideas of consciousness
one must turn to Aurobindo (e.g. 1939, 1956).

\section{Concluding Remarks}

It appears that Indian  understanding of physics was
informed not only by astronomy and terrestrial experiments
but also by speculative thought and by
meditations on the nature of consciousness.
Unfettered by 
either geocentric or anthropocentric
views, this understanding unified the physics of
the small with that of the large within a framework that
included metaphysics.

This was a framework consisting of innumerable worlds
(solar systems), where time and space were continuous,
matter was atomic, and consciousness was atomic, yet
derived from an all-pervasive unity.
The material atoms were defined first by
their subtle form, called {\em tanm\={a}tra},
which was visualized as a potential, from
which emerged the gross atoms.
A central notion in this system was that all
descriptions of reality are circumscribed by paradox
(Kak, 1986).

The universe was seen as dynamic, going through ceaseless
change.

\section{References}
\begin{description}
 
\item Sri Aurobindo, 1939. {\em The Life Divine}. Aurobindo Ashram, 
Pondicherry.
 
\item Sri Aurobindo, 1956. {\em The Secret of the Veda}. Aurobindo Ashram, 
Pondicherry.

\item C. Chapple, 1984. Introduction and bibliography in Venkatesananda (1984).
 
\item S. Dasgupta, 1975. {\it A History of Indian Philosophy.} Motilal
Banarsidass, Delhi.

\item G. Feuerstein, S. Kak, D. Frawley, 1995. {\em In Search of the Cradle
of Civilization.} Quest Books, Wheaton.

\item K.M. Ganguly (tr.), 1883-1896. {\em The Mah\={a}bh\={a}rata.}
Reprinted Munshiram Manoharlal, Delhi, 1970.

\item P. Halpern, 1995. {\em The Cyclical Serpent: Prospects for an
Ever-Repeating Universe.} Plenum Press, New York.

\item S. Kak, 1986. {\it The Nature of Physical Reality.}
Peter Lang, New York.

\item S. Kak, 1990. Kalidasa and the Agnimitra problem.
{\em Journal of the Oriental Institute} 40: 51-54.

\item S. Kak, 1994. {\it The Astronomical Code of the \d{R}gveda.}
Aditya, New Delhi.

\item S. Kak, 1995a. From Vedic science to Ved\={a}nta. 
{\it Brahmavidy\={a}:
The Adyar Library Bulletin}, 59: 1-36.

\item S. Kak, 1995b. The astronomy of the age of geometric altars. {\em Quarterly
Journal of the Royal Astronomical Society} 36: 385-396.

\item S. Kak, 1996. Knowledge of planets in the third millennium BC.
{\em Quarterly
Journal of the Royal Astronomical Society} 37: 709-715.

\item S. Kak, 1997. On the science of consciousness in ancient India.
{\em Indian Journal of History of Science}  32: 105-120.

\item S. Kak, 1997-8. Vai\d{s}\d{n}ava metaphysics or a science of consciousness.
{\em Pr\={a}chya Pratibh\={a}} 19: 113-141.

\item S. Kak, 1997-8. Consciousness and freedom according to the \'{S}ivaS\={u}tra.
{\em Pr\={a}chya Pratibh\={a}} 19: 233-248.

\item S. Kak, 1998a. The speed of light and Pur\={a}\d{n}ic cosmology. LANL physics archive
9804020. Also in Rao and Kak (1998).

\item S. Kak, 1998b. S\={a}ya\d{n}a's astronomy. {\em Indian Journal of History of
Science} 33: 31-36.
 
\item S. Kak, 1998c. Early theories on the distance to the sun.
{\em Indian Journal of History of Science} 33: 93-100.

\item S. Kak, 1998d. The orbit of the sun in the Br\={a}hma\d{n}as. 
{\em Indian Journal of History of Science} 33: 175-191.

\item S. Kak, 1999. Physical concepts in S\={a}\d{m}khya and Vai\'{s}e\d{s}ika. Chapter in
{\it Science and Civilization in India, Vol. 1, Part 2}, edited by G.C. Pande,
Oxford University Press, Delhi, in press.

\item W. Moore, 1989. {\it Schr\"{o}dinger: Life and Thought.} Cambridge
University Press, Cambridge.

\item T.R.N. Rao and S. Kak, 1998. {\it Computing Science in Ancient
India.} USL Press, Lafayette.


\item E.C. Sachau, 1910. {\em Alberuni's India.} Reprinted by Low Price Publications,
Delhi, 1989.

\item E. Schr\"{o}dinger, 1961. {\it Meine Weltansicht.} Paul Zsolnay, Vienna.

\item F. Staal, 1988. {\it Universals.} University of Chicago Press, Chicago.
 
\item S. Venkatesananda (tr.), 1984. {\it The Concise Yoga V\={a}si\d{s}\d{t}ha.} State University of New York Press, Albany.

\item S. Venkatesananda (tr.), 1993. {\it V\={a}si\d{s}\d{t}ha's Yoga.}
State University of New York Press, Albany.

\item {\it Yoga V\={a}si\d{s}\d{t}ha}, 1981. Munshiram Manoharlal, Delhi.

\item V.S. Wakankar, 1992. Rock painting in India. In
{\it Rock Art in the Old World,} M. Lorblanchet (ed.). 319-336.
New Delhi.
 
\end{description}
 
\end{document}